\documentclass[fleqn,10pt]{wlscirep}
\usepackage[utf8]{inputenc}
\usepackage[T1]{fontenc}

\usepackage{graphicx}
\usepackage{hyperref} 

\def \bp{$\beta$\,Pic}
\def \betapictoris{$\beta$\,Pictoris}
\def \alphapic{$\alpha$\,Pic}
\def \ds{$\delta$ Scuti}
\def \sun{\ensuremath{\odot}}

\title{Exocomets size distribution in the \betapictoris\ planetary system}

\author[1,*]{Alain Lecavelier des Etangs}
\author[1,2]{Lucie Cros} 
\author[1,3]{Guillaume H\'ebrard}
\author[1,4]{Eder Martioli}
\author[5]{Marc~Duquesnoy}
\author[6]{Matthew Kenworthy}
\author[1,5]{Flavien Kiefer}
\author[5]{Sylvestre Lacour} 
\author[5]{Anne-Marie~Lagrange}
\author[7]{Nad\`ege Meunier} 
\author[1]{Alfred Vidal-Madjar}

\affil[1]{Institut d'Astrophysique de Paris, CNRS, UMR 7095, Sorbonne Universit\'e, 98 bis bd Arago, 75014 Paris, France}
\affil[2]{Ecole nationale sup\'erieure des mines de Paris, Universit\'e PSL, 60 boulevard Saint-Michel, 75272 Paris, France}
\affil[3]{Observatoire de Haute-Provence, St Michel l'Observatoire, France}
\affil[4]{Laborat\'{o}rio Nacional de Astrof\'{i}sica, Rua Estados Unidos 154, 37504-364, Itajub\'{a} - MG, Brazil}
\affil[5]{LESIA, Observatoire de Paris, Universit\'e PSL, CNRS, Sorbonne
Universit\'e, Univ. Paris Diderot, Sorbonne Paris Cit\'e, 5 place Jules
Janssen, 92195 Meudon, France}
\affil[6]{Leiden Observatory, University of Leiden, The Netherlands}
\affil[7]{Univ. Grenoble Alpes, CNRS, IPAG, 38000 Grenoble, France}

\affil[*]{lecaveli@iap.fr\vspace{1cm}}

\affil[**]{
{\bf Note} : This is a preprint of an article published in Scientific Reports (Springer Nature).\newline
The final version is available online from April 28, 2002 at:
\href{https://www.nature.com/articles/s41598-022-09021-2}
{https://www.nature.com/articles/s41598-022-09021-2}
}

\begin{abstract}

The star \betapictoris\ harbors a young planetary system, which is characterized by the presence of a gaseous and dusty debris disk \cite{Smith_1984, Vidal-Majar_1986, Kalas_2001, Roberge_2006,Apai_2015,Brandeker_2016}, at least two massive planets \cite{Lagrange_2010,Snellen_2018, Lagrange_2019,Lacour_2021} and many minor bodies. For more than thirty years, exocomets transiting the star have been detected using spectroscopy, probing 
the gaseous part of the cometary comas and tails
\cite{Ferlet_1987,Beust_1990,Vidal-Madjar_1994,Kiefer_2014,Strom_2020}. \\
The detection of the dusty component of the tails can be performed through photometric observations of the transits\cite{Lecavelier_etal_1999, Lecavelier_1999, Rappaport_2018, Zieba_2019}. 
Since 2018, the Transiting Exoplanet Survey Satellite\cite{Ricker_2015} has observed \bp\ for a total of 156~days. 
Here we report an analysis of the TESS photometric data set with the identification of a total of 30 transits of exocomets. 
Our statistical analysis shows that the number of transiting exocomet events ($N$) 
as a function of the absorption depth ($AD$) in the light curve follows a power law in the form 
$dN(AD) \propto AD^{-\alpha}$, 
where $\alpha=2.3\pm 0.4$. 
This distribution of absorption depth leads to a differential comet size distribution proportional to 
$R^{-\gamma}$, where $\gamma =3.6 \pm 0.8$, 
showing a striking similarity 
to the size distribution of comets in the Solar system and the distribution of a collisionally relaxed population ($\gamma_{\rm D}= 3.5$). 

\end{abstract}
\begin{document}

\flushbottom
\maketitle
\thispagestyle{empty}

\newpage

\section*{Introduction}
\label{sec:Introduction}

Since the mid-1980's, spectroscopic observations of the bright southern star \betapictoris , or \bp , have revealed variations in the calcium H and K lines, which have been interpreted as the transit 
of the gaseous tails of exocomets 
\cite{Ferlet_1987,Beust_1990,Vidal-Madjar_1994,Kiefer_2014}. 
Using data obtained by the Transiting Exoplanet Survey Satellite (TESS\cite{Ricker_2015})
between October 2018 and February 2019, three 
photometric events were discovered \cite{Zieba_2019} and interpreted as due to the transit of the dust component of exocomets. In support of this interpretation, the observed light curves are almost identical 
to the predictions made twenty years earlier \cite{Lecavelier_etal_1999, Lecavelier_1999}.

Since 2019, \bp\ has been re-observed by TESS. 
Here we present the analysis of the complete data set gathered up to February 2021 in order to perform a deep search for exocomet transits and determine 
the size distribution of the \bp\ comets to be compared with the distribution observed in the Solar system.

\section*{TESS observations}
\label{sec:TESS observations}

\betapictoris\ has been observed by TESS at 2-minute cadence several times from October 2018 to February 2021.
The available data-set covers a total of 156~days of observations in the optical domain, 
divided into 14 epochs of about 12 days each. 
The raw data shows a flux dispersion of about $10^{-3}$, which is mainly due to \ds\ pulsations in the stellar atmosphere.

The data from 19 October 2018 to 1 February 2019 have already been analysed\cite{Zieba_2019}. Three photometric events have been identified and attributed to the transits of three different exocomets, with one spectacular transit at Julian Day (JD) equals to (2457000+1486).

In addition to these pioneering observations, \bp\ has been observed from 20 November 2020 to 8 February 2021. We downloaded the whole data set from the TESS database at MAST for the observations of Sectors 4 to 7 and Sectors 31 to 34 (Extended Data Table~\ref{tab:tess_data}).

\section*{Search for exocomet transits}
\label{sec:Search for exocomets transits}

After the cleaning process of the \ds\ variations and other slower variations (see Method), we end up with a 156~days light curve clearly showing some dips, which are similar to what is expected for exocomet transits (Extended Data Fig.~\ref{fig:dips}). 
To make sure that the observed dips in the \bp\ light curve are real and not noise residuals nor artifacts due to the reduction process, 
we downloaded the TESS observations of the nearby star \alphapic\ and applied the same procedure to remove the \ds\ and slower variations. 
The observations of \alphapic\ provide an excellent data-set to test our procedure 
because \alphapic\ is in the same region of the sky as \bp\ 
(and has hence overlapping TESS observations epochs), 
it has the same spectral type (A8V versus A5V for \bp ), and similar magnitude (3.3 versus 3.85 for \bp ). 
Thus, \alphapic\ is almost a nearby stellar twin of \bp\ except for the presence of the young planetary system. 
\alphapic\ has already been successfully used as a reference star for analysis of \bp\ observations \cite{Lecavelier_1993}.   

We first compare \alphapic\ and \bp\ by visual inspection of plots. The light curve of \alphapic\ shows noisy excursions from the mean value both in the positive and negative deviations with the same pattern. 
The light curve of \bp\ also shows noisy excursions in the positive direction that are similar to the ones observed in the light curve of \alphapic , but noisy excursions that are more pronounced and more frequent in the negative direction. The latest are typical signatures of small, transiting objects with extended dust tails.

We characterize this excess of transit-like features in the light curve of \bp\ and identify the corresponding individual events. To do so, we calculated the correlation of the light curves with a simple model of an exocomet transit photometric event, assuming a 1D transit of a translucent dust cloud with an exponential decrease of the optical thickness from the head of the comet. 
This model has only four parameters : 
$K$, the cloud optical thickness at the leading head, 
$\Delta t$, the transit duration 
corresponding to the time needed to cover 
the chord length of the stellar disk at the transit velocity, 
$\beta$, the speed of the transit of one scale length of the cometary tail (the inverse of the scale length of the cometary tail divided by the transit velocity), 
and finally $t_0$, the time of the beginning of the transit.
With this model of the transit of the exocometary tail, 
the corresponding decrease in relative flux at the time $t$ is given by 
$\Delta F/F\, (t)  = K \times \left(\exp (-\Delta) - \exp (-\Delta ^\prime )\right) $, 
where $\Delta=\beta (t-t_0)$ if $t\ge t_0$, 
$\Delta=0$ if $t\le t_0$, 
and $\Delta^\prime =\beta (t-t_0-\Delta t)$ if $t\ge t_0 +\Delta t$,
and $\Delta^\prime =0$ if $t \le t_0 +\Delta t$.

We calculated the correlation of this 1D-model with the observed light curves of \bp\ and \alphapic\ 
by varying the value of the transit time $t_0$ and using various plausible values for the $\Delta t$ and $\beta$ parameters characterizing the shapes of the exocomet transit light curves. 
We used $\beta$ from 2 to 20\,days$^{-1}$ 
and $\Delta t$ from 0.15 to 0.5\,days, corresponding to periastron distances ranging from about 0.08 to 0.85\,au.
The value of $K$ in the model can be arbitrarily chosen because it only changes the amplitude of the light variations, hence it has no consequence in the position of the correlation maximum nor on the identification of exocomet transits events.  

In the case of \bp\ the correlation reaches large positive values for 
some of the values of $t_0$. This behaviour is not observed with the \alphapic\ light curve, showing that the photometric variations
with cometary transit shapes are specific to \bp . 
We check the negative values of the correlation of the model with the \bp\ light curve itself, 
and found that they are much less numerous and significantly smaller than the positive values. This confirms that the 
light curve of \bp\ shows photometric variations with decrease of the star brightness that are typical of the transits of exocomets and the absence of variations with increase of the star brightness with similar shape and  amplitude. 
We interpret the correlation peaks as the signatures of potential exocomet transits. 
We keep only the peaks with correlation values that are higher than the maximum value obtained with the \alphapic\ data and the maximum negative value with the \bp\ data. 
This is our conservative criterion to consider the observed 
variations as a detection of an exocomet transit. 

We identified a total of 30~significant detections of exocomet transit events. 
This confirms the ubiquity of comets in that young planetary system.
The light curves of these detected exocomet transits are plotted in Extended Data Fig.~\ref{fig:dips}. 
Their characteristics and the parameters of the best fits with the 1-D model are given in Extended Data Table~\ref{tab:List detections}.

\section*{Size distribution of $\beta$\,Pictoris exocomets}
\label{sec:Size distribution of beta Pictoris exocomets}

\subsection*{Distribution of absorption depths}
 \label{sec:Distribution of absorption depths}
 
\begin{figure}[bp!]
   \centering
   \includegraphics[width=1\hsize]{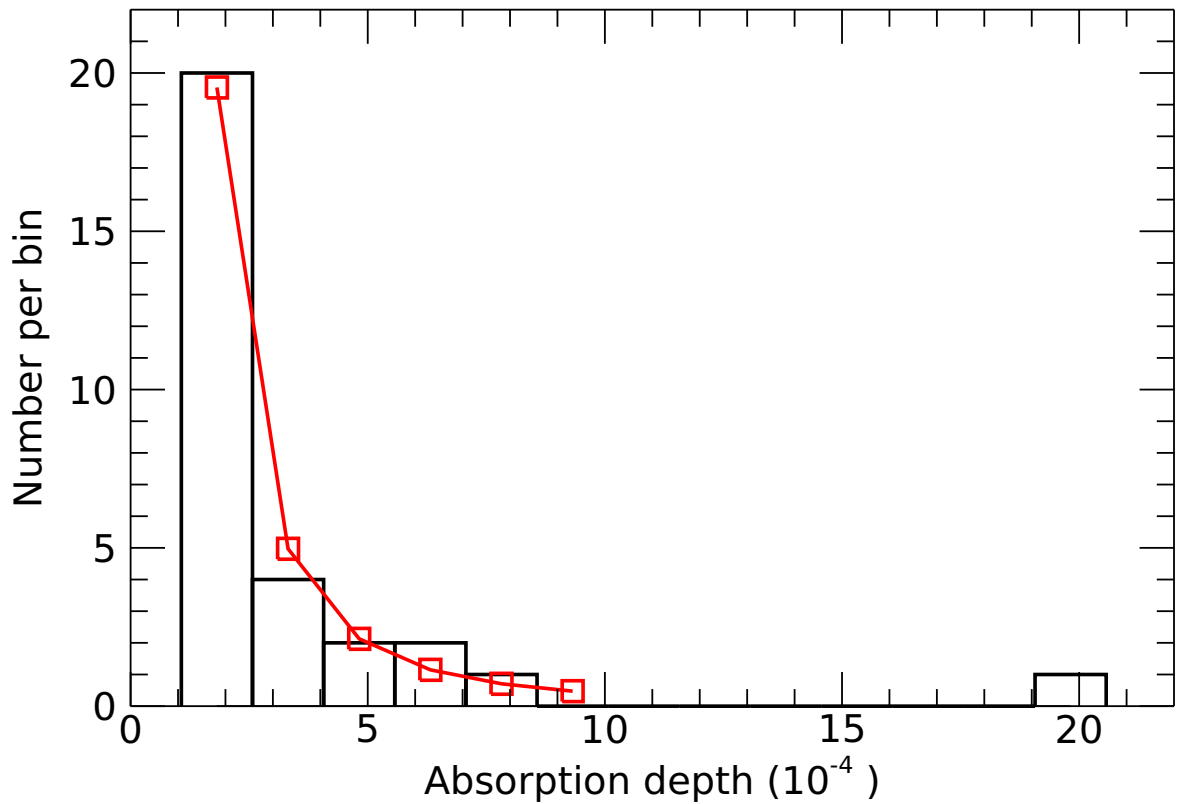}
      \caption{{\bf Histogram of the number of exocomet transit events as a function of the absorption depth}. The red line shows the fit with a power law function with $\alpha=2.3 \pm 0.4$. The uncertainty on the fitted parameters of the power law function have been evaluated using a Poisson distribution for the number of events in each bin of width $d AD=1.5\cdot 10^{-4}$. 
      The red squares represents the number of expected events in each bin as calculated with the fitted power law.}
        \label{fig:histogram}
  \end{figure}

Our detection of 30~photometric transits of exocomets 
allows a statistical analysis of their properties.
Here we call "absorption depth", noted $AD$, 
the decrease in relative flux at the minimum of a transit light curve. The numerical value of the absorption depth is estimated from the best fit with the 1-D model with  $AD=K(1-\exp(-\beta \Delta t))$. 
A plot of the events frequency as a function of the absorption depth shows that there is a steep decrease of the number of events toward the larger absorption depths
(Fig.~\ref{fig:histogram}). 
The differential number of transiting exocomet events ($dN$) detected with an observation of duration $\delta t$ as a function of the absorption depth can be fitted by a power law in the form $dN(AD)=N_{0}\cdot (AD/10^{-4})^{-\alpha}\cdot (\delta t/100\, {\rm days})\cdot (dAD/10^{-4})$. 
Considering 29~events detected in 156~days of observations,
that is all the 30~events except the deepest one (see below), 
we find that 
 $\alpha=2.3 \pm 0.4$ and $N_{0}=33^{+16}_{-11}$, 
 where the uncertainties have been evaluated 
 using a Poisson distribution for the number of events 
 in each bin of width $d AD=1.5\cdot 10^{-4}$.  
 
The deepest transit event of Julian Day JD=2457000+1486 looks exceptional with an absorption depth of about $20\times 10^{-4}$. This event could be produced by a member of another family of exocomets than the one which produces the 29~other shallower events, as we know from spectroscopic transit observations the presence of several families of \bp\ comets \cite{Kiefer_2014}. 
Nonetheless, with the distribution derived above, 
the expected number of events with absorption depths in the range [10-20]$\times 10^{-4}$ in a 156~days observation 
is $1.2^{+0.6}_{-0.4}$.
Even in the range [15-20]$\times 10^{-4}$ the expected number of detections in 156~days 
is $0.38^{+0.19}_{-0.12}$, corresponding to a probability of 26\% to have one single event in this range as observed. Therefore the event of JD=2457000+1486 can simply be a normal event, only the deepest, within the same distribution of the other 29~events. Nonetheless, to remain conservative, this deepest event is not taken into account in the derivation of the distribution of absorption depths and exocomets sizes.

\subsection*{Distribution of exocomet sizes}

The modeling of the exocomet transit light curves shows that 
the transit absorption depth, $AD$, is directly proportional 
to $\dot{M}$, the dust evaporation rate from the comet nucleus\cite{Lecavelier_etal_1999} . 
If we assume that the dust production rate is proportional to the comet nucleus area ({\it i.e.}, ref.\cite{Jewitt_1999}), 
we have a production rate $\dot{M}$ proportional to the squared of the nucleus radius $R$.
Finally, with $AD$ proportional to $R^2$, 
we find that the differential number of exocomets as a function of the nucleus size is given by 
$dN(R) \propto R^{-\gamma} dR $, with $\gamma = 2\alpha-1$. 

\begin{figure*}[p!]
   \centering
   \includegraphics[width=\textwidth]{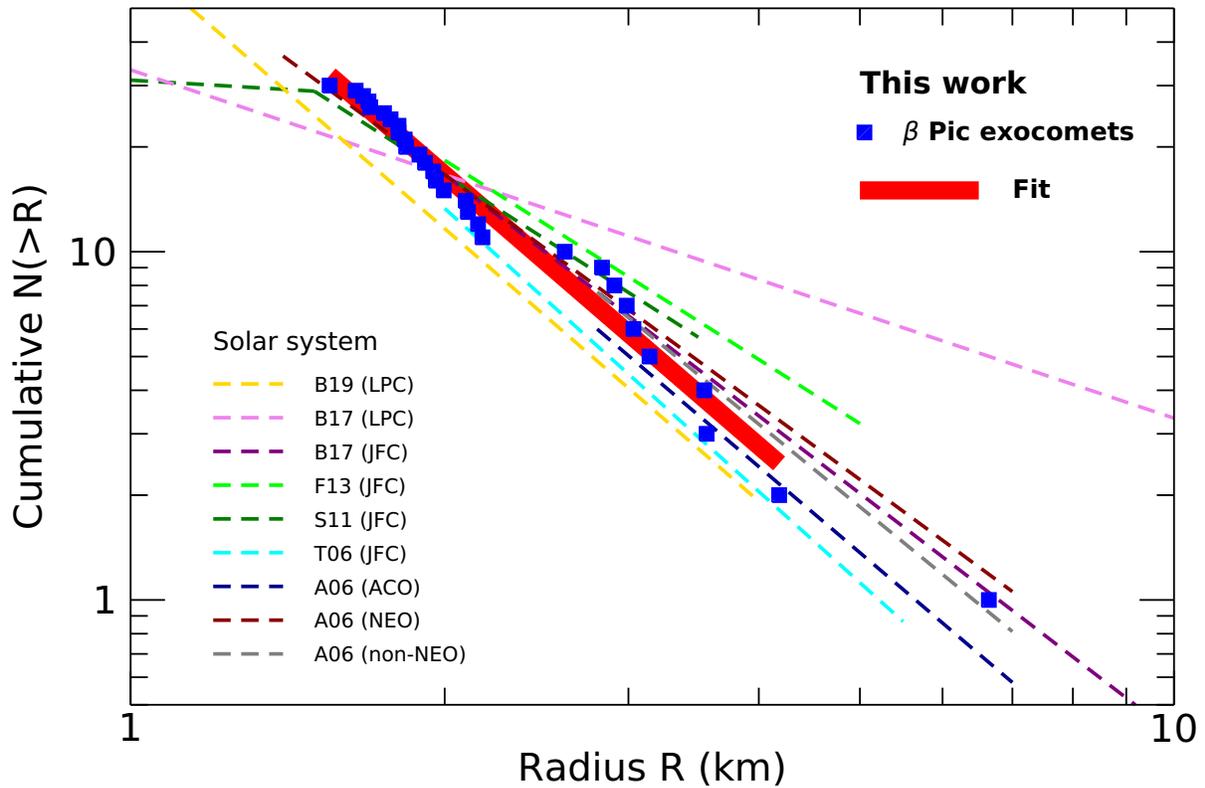}
      \caption[]{{\bf Plot of the cumulative size distribution of the exocomets
      in \bp .} The cumulative size distribution is plotted with blue squares for each exocomet and the corresponding fit 
      excluding the largest comet is plotted with the red thick line. 
      For comparison, published size distributions measured 
      in the Solar system are plotted with thin dashed lines 
      for asteroids in comets orbits (ACO), on near Earth orbits (NEO) 
      and non-near Earth orbits (non-NEO) (A06, ref.\cite{Alvarez_2006}),  
      Jupiter family comets (JFC) (T06, ref.\cite{Tancredi_2006} ; 
      S11, ref.\cite{Snodgrass_2011} ; F13, ref.\cite{Fernandez_2013} ; 
      B17, ref.\cite{Bauer_2017}), 
      and long-period comets (LPC) (B17, ref.\cite{Bauer_2017} ; 
      B19, ref.\cite{Boe_2019}).
      In this plot, the size distributions for the objects in the Solar system 
      have been scaled to have a cumulative number of about 10 objects with radius above 2~kilometers.
      
      The radii of the \bp\ comets have been estimated using the derivation described in the Method section.
      The conclusion on the similarity of the size distributions in \bp\ and the Solar system is independent of these estimates.}
        \label{fig:Radius_Cumulative_Distribution}. 
  \end{figure*}

With the fit to the observed distribution of the absorption depths, 
we conclude that the differential distribution of the exocomet size must follow a power law with an index $\gamma =3.6 \pm 0.8$. 
This distribution is notably similar to the size distribution of comets in the Solar system (Fig.~\ref{fig:Radius_Cumulative_Distribution}) 
and the distribution predicted in ref.\cite{Dohnanyi_1969} for a collisionally relaxed population
($\gamma_{\rm D}= 3.5$).

For the plot of the size distribution (Fig.~\ref{fig:Radius_Cumulative_Distribution}), we used the cometary radii estimated following the derivation described in the Method section. 
The conclusion on the similarity of the size distributions in \bp\ and the Solar system is independent of these absolute size estimates. 
Nonetheless, it is remarkable that not only the distribution but also the sizes of the \bp\ comets nuclei are found to be similar to sizes of the Solar system comets.

\subsection*{Comparison with Solar system comets}

The size distribution of the nucleus of Solar system comets has been estimated for various locations. The size distribution of the Jupiter family comets and Oort cloud comets are found to be similar but not exactly the same.

For the Oort cloud, the size distribution has been estimated using a cometary activity model with a survey simulation 
and application to 150~long-period comets (LPC) detected over 7~years by the Pan-STARRS1 near-Earth object survey \cite{Boe_2019}. 
For objects with diameters above 1~km, the distribution is found to be $\gamma = 3.6\pm 0.4$\,(stat.)\,$\pm  0.7$\,(sys.), 
which is in remarkable agreement with our value for \bp\ exocomets. 
For smaller long-period comets, a shallower distribution is found
with $\gamma = 0.35$ for diameters between 100~m and 1~km (ref.\cite{Boe_2019}). 
Note that in practice, in a similar manner as we have used the transit absorption depth 
as a proxy for the size estimate of the \bp\ comets, 
the above estimates for the Oort cloud comets have been obtained using the absolute H magnitudes of the nuclei as the proxy for their size. 

For the \bp\ exocomets, the comparison may be more appropriate 
with the size distribution of the comets in the Jupiter family, 
which originates from the Kuiper Belt. 
In ref.\cite{Tancredi_2006} a catalog of absolute nuclear magnitudes 
of Jupiter family comets (JFC) has been used to derive a size distribution 
and to find $\gamma=3.7 \pm 0.3$ for nuclei 
with radius between 2 and 5.5~kilometers (see Fig.~8 of ref.\cite{Tancredi_2006}).
More recent works have provided a shallower distribution :   
analyzing a large number of optical observations, 
ref.\cite{Snodgrass_2011} found a lower value with $\gamma=2.9 \pm 0.2$ 
for nuclei with radius larger than 1.25~kilometers,
and $\gamma = 1.2$ for smaller objects.
This last result is consistent with the result described in ref.\cite{Meech_2004}, 
where images of Jupiter family comets 
obtained with the Hubble space telescope and the Keck telescopes have been analyzed. 
With a model fit to the observations, it is  
concluded that the intrinsic size distribution of comets in the Jupiter family 
is consistent with a $\gamma = 3.5$ power-law 
but truncated at small nucleus radii below 2.0~kilometers.
In ref.\cite{Fernandez_2013} a similar distribution is obtained with $\gamma = 2.9$ 
for radius between 2 and 5~kilometers, 
interpreting the distribution shallower than the canonical Dohnanyi’s size distribution 
 $\gamma_{\rm D} = 3.5$ (ref.\cite{Dohnanyi_1969}) as due to fragmentation of the JFC objects.
In ref.\cite{Bauer_2017} the measured $\gamma= 3.3 \pm 0.2$ 
for Jupiter family comets of radius between 2 and 10~kilometers 
is to be compared to the $\gamma =2.0 \pm 0.1$ for long-period comets
between 1 and 20~kilometers in radius.

Finally, in the Solar system the size distribution of extinct or dormant comets 
can be determined through the population of asteroids in comet orbits (ACO). 
For these objects, ref.\cite{Alvarez_2006} found $\gamma=3.55 \pm 0.04$ for the full sample with radius between 2.8 and 7~kilometers, $\gamma=3.2 \pm 0.04$ for near Earth objects (NEO) with radius down to 1.4~kilometers and $\gamma=3.45 \pm 0.04$ for non-near Earth objects (non-NEO) with radius between 2.8 and 7~kilometers. 

Taken all together these estimates for the Solar system comets are in general agreement 
with the value that we obtained for the \bp\ comets, with some slightly shallower distributions in some cases (Fig.~\ref{fig:Radius_Cumulative_Distribution}). This points toward the importance of collisional fragmentation in shaping the
size distribution of the exocomets in the younger \bp\ planetary system.

\section*{Discussion}

The measured absorption depth distribution is the result of the distribution of several parameters for each individual comet, {\it e.g.}, the orbital parameters, cometary activity, size, etc. Here we assumed that the absorption depth distribution is mainly dominated by the distribution of the exocomet intrinsic dust production rate and hence their size. 
In other terms, although other parameters play a role for each individual comet, their diversities are expected to have a lower impact on the observed transit absorption depth than the size. In support of that idea, in spectroscopy it is observed that the \bp\ transiting comets present similar orbital characteristics, which allows the classification in two different families \cite{Kiefer_2014}. With similar orbits, different transiting exocomets have different dust tails mainly because of different dust production rate, and hence because of different size nuclei. 

The observed distribution of exocomets in the young planetary system of \bp\ 
is strikingly similar to the distribution observed in the Solar system. 
This distribution seems to be ubiquitous 
and is also consistent with the canonical Dohnanyi's size 
distribution \cite{Dohnanyi_1969} ($\gamma_{\rm D} = 3.5$), 
which corresponds to the size distribution of a collisionally relaxed population (see discussion in ref.\cite{OBrien_2005}). 
This indicates that the collisional process with fragmentation cascades
is likely one of the dominant processes that shape 
the population of kilometer-sized bodies in the \bp\ planetary systems.

\newpage

\section*{Methods}

\subsection*{Cleaning the light curves}
\label{sec:Cleaning the light curves}

\subsubsection*{The \ds\ variations}
\label{sec: The delta sculti variations}

$\beta$\,Pictoris is prone to \ds\ type photometric variations \cite{Koen_2003,Mekarnia_2017,Zwintz_2019}. 
These variations have a dominant frequency of 47.44\,d$^{-1}$ (corresponding to a period of about 0.5~hours), with an amplitude of up to $4\times 10^{-3}$ (Extended Data Fig.~\ref{fig:delta scuti variations}).
These variations are superimposed on the searched signal and must be corrected before looking for exocomet transit signatures. 

Each of the 14 epochs of continuous observations (Extended Data Table~\ref{tab:tess_data}) have been reduced separately ; for each of them we extracted the set of frequencies and amplitudes of the pulsations using the {\tt Period04} software as previously done in ref.\cite{Zieba_2019}. The {\tt Period04} software performs a Fourier transform on the data, giving the frequencies, amplitude, phase and signal to noise ratio of each harmonic in the time series. 
We conducted the frequency search between 15 and 100 day$^{-1}$, to avoid removing slow variations that might be caused by a cometary transit. 
For each iteration of the software, the highest-amplitude frequency within the search range is selected, and added to a multi-sine model. 
This model is then optimized over amplitude and frequency of each harmonic, then removed from the original signal. 
For each iteration, the software computes the signal to noise ratio of the main remaining frequencies. 
The exit condition of the loop was chosen when the main frequency's signal to noise ratio went under~4. 
We considered that below this limit, all that is left is noise.

The process has been applied for each of the 14~blocks of continuous observations, leading to an average number of 43~different pulsations for each block.
After identifying all these pulsations, we subtracted them from the photometric measurements. The resulting residuals were then rebinned from an initial time sampling of 120~seconds to a time sampling of 1800~seconds.

To validate the result, we checked that the three exocomet transits already identified in ref.\cite{Zieba_2019} are clearly visible in the data set cleaned from the \ds\ variations by this procedure (Extended Data Fig.~\ref{fig:clean light curve}). 
The data are clean enough that new potential exocomet transits are suspected from the resulting light curve.

\subsubsection*{Remaining slow variations}

After removal of the \ds\ variations, the resulting light curve still shows 
slow variations. 
These variations can have various origins (residuals of the \ds\ variations that were not properly eliminated, systematic correlated noise, instrumental effects, etc.). 
Whatever the origin of these slow variations, astrophysical or instrumental, they need to be cleaned before searching for exocomet transits. 
To do so, for each epoch of continuous observations we searched for a smooth function modelling these variations. 
First we rebinned the time series at a 1-day-long interval, in order to "protect" the shorter dimmings (such as comets), excluded the 1-day measurements where the deepest exocometary transits are already clearly identified (by using the detection procedure described in the main text),
then we interpolated a function with a cubic spline to model the generic form of the time series.
Finally, we normalized our time series with this model. The result is a flattened light curve, which can be directly used to search for shallow dip events due to exocomet transits.
The procedure is illustrated in Extended Data Fig.~\ref{fig:final light curve} where a potential second shallower exocomet transit closely follows another deep exocomet transit event.

\subsection*{Comet's nuclei radius}
\label{Appendix:radius}

The conclusion on the size distribution given in the main text does not require to estimate the true size of the comet's nuclei. It only relies on the assumption that the absorption depth is a good proxy for the dust production rate, which is supported by numerical simulations of exocomets light curves \cite{Lecavelier_1999}, and that the production rate is proportional to the area of the comet's nucleus, which is consistent with Solar system comets' models and observations.

Nonetheless, using these simulations and Solar system observations, we can make a step further and derive the typical sizes of the comets detected in the \bp\ TESS light curve. From a newly calculated library of exocomet transit light curves similar to the one of ref.\cite{Lecavelier_1999}, we derive a typical scaling law for the absorption depth $AD$, which is 
$$
AD\simeq 5\cdot 10^{-5}\left(\frac
{\dot{M_{1\,au}}}{10^5 kg\, s^{-1}}
\right)
\left(\frac{q}{1\,au}\right)^{-1/2}
\left(\frac{M_*}{M_\sun}\right),
$$
where $\dot{M_{1\,au}}$ is the dust production rate of the comet when it is at 1\,au from the star, $q$ is the orbital periastron distance and $M_*$ is the mass of the star. This estimate is valid over a wide range of the longitude of perisastron of the comet's orbit, $\omega$.

The periastron distance of the detected comets can be estimated using the transit time $\Delta t$. 
We have $\Delta t = L_{chord}/v_{\rm transit}$, 
where the mean chord length is $\overline{L_{chord}}=\pi  R_{*}/2$ 
and the transit velocity is 
$v_{\rm transit}=\sqrt{GM_* /2q}(\cos \omega +1)\sim \sqrt{GM_*/q}$.
With a \bp\ radius of $R_*=1.7 R_{\sun}$ (ref.\cite{Kervalla_2004}) and a mass of $M_*=1.75 M_{\sun}$, we find $\Delta t \simeq 13 (\sqrt{q/1\,au}$)\,hours. The best fits values of $\Delta t$ correspond to distances ranging from 0.03 to 1.3\,au, in good agreement for the distances expected for the comet evaporation.
The mean value of the estimated periastron distances is about 0.18\,au. Using this mean distance, we obtain the following relationship for the observed absorption depth and the dust production rate :
$$
AD\simeq 2\cdot 10^{-4} 
\left(\frac{\dot{M_{1\,au}}}{10^5 kg\, s^{-1}}\right).
$$

Finally, the relation between the evaporation rate and the comet's nucleus size can be derived by scaling the observation of the Hale-Bopp comet. 
Using a radius of about 30\,kilometers \cite{Jewitt_1999,Fernandez_1999, Bair_2018} 
and a dust production rate of 
$2\cdot 10^6$\,kg\,s$^{-1}$ at 1\,au (ref.\cite{Jewitt_1999}) 
for this well-observed dusty comet, we find 
$\dot{M_{1\,au}}\simeq 2\cdot 10^6$\,kg\,s$^{-1}$
$(R/30$\,km$)^2(L_*/L_{\sun})$.
With a \bp\ luminosity of 8.7$L_{\sun}$, we find
$$
R\simeq 7.2 \, {\rm km} 
\left(\frac{\dot{M_{1\,au}}}{10^6\,{\rm kg \, s}^{-1}}\right)^{1/2}.
$$

All together, we conclude that the radius of the \bp\ comets nuclei can be estimated using the photometric transit absorption depth with
$$
R\simeq 1.5 {\rm km} \sqrt{AD/10^{-4}}.
$$

Using this relationship, we derive a size of 1.5\,km for the smallest detected comets ($AD\simeq 10^{-4}$), and 6.7\,km for the largest comet ($AD\simeq 20\cdot 10^{-4}$). These sizes are remarkably similar to the sizes of comets in the Solar system.

\bibliography{bibliography}

\section*{Acknowledgements}

We thank Paul Wiegert for enlightening discussion on the size distribution of comets in the Solar system. This paper includes data collected with the TESS mission, obtained from the MAST data archive at the Space Telescope Science Institute (STScI). 
Funding for the TESS mission is provided by the NASA Explorer Program. STScI is operated by the Association of Universities for Research in Astronomy, Inc., under NASA contract NAS 5–26555. We thank the TESS Team members for making available the extremely accurate photometric data they obtained. 
ALdE, LC, GH, EM, FK, and AVM acknowledge support from the CNES (Centre national d'\'etudes spatiales, France).

\section*{Author contributions statement}

ALdE initiated and directed the project, analysed the cleaned light curves, interpreted the result and wrote the bulk of the manuscript. 
LC analysed the raw TESS data, developed and applied the algorithms, produced the cleaned light curves, 
extracted the first list of exocomet transits, participated to the interpretation and wrote parts of the manuscript. 
GH contributed to the project design, definition of the scientific goal, the derivation of Fig. 2. and to the writing. 
EM contributed to the TESS data extraction and interpretation, and to the writing. 
MD, AML and NM developed an independent analysis of the data for a cross-check of the result, 
extracted a confirmation list of exocomet transits, and participated to the writing.
MK, FK, SL and AVM participated to the definition of the project, scientific interpretation of the results, 
clarification of the presentation and to the writing. 
All authors discussed the scientific result, edited the manuscript and contributed to the final version. 

\section*{Data availability}

The observational data used in this work are publicly available in the Mikulski Archive for Space Telescope (MAST). 
The data in the tables and the final cleaned light curves are publicly available on GitHub at https://github.com/lecaveli/BetaPic\_TESS.

\section*{Additional information}

\subsection*{Competing interests} 
The author declare no competing interests.

\newpage

\section*{Extended Data Figures}

\setcounter{figure}{0}
\renewcommand{\figurename}{Extended Data Fig.}

  \begin{figure}[htp!]
   \centering
   \includegraphics[width=1\hsize]{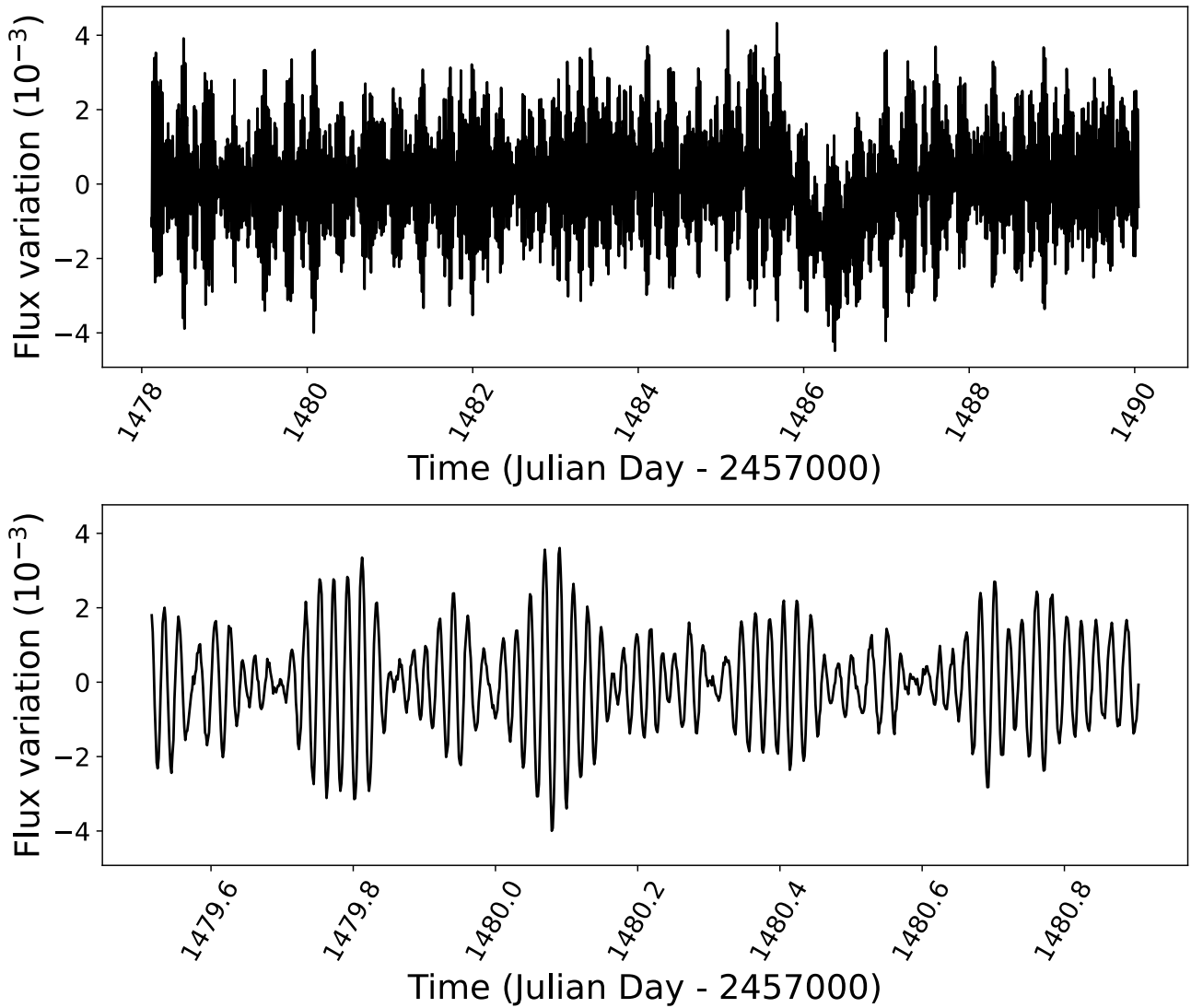}
      \caption{{\bf Sample of TESS photometry of \bp .} The light curve is dominated by \ds\ variations. The bottom panel shows the variations over about half a day with a relative amplitude of up to $4\times 10^{-3}$.}
        \label{fig:delta scuti variations}
  \end{figure}

  \begin{figure}[htp!]
   \centering
   \includegraphics[width=1\hsize]{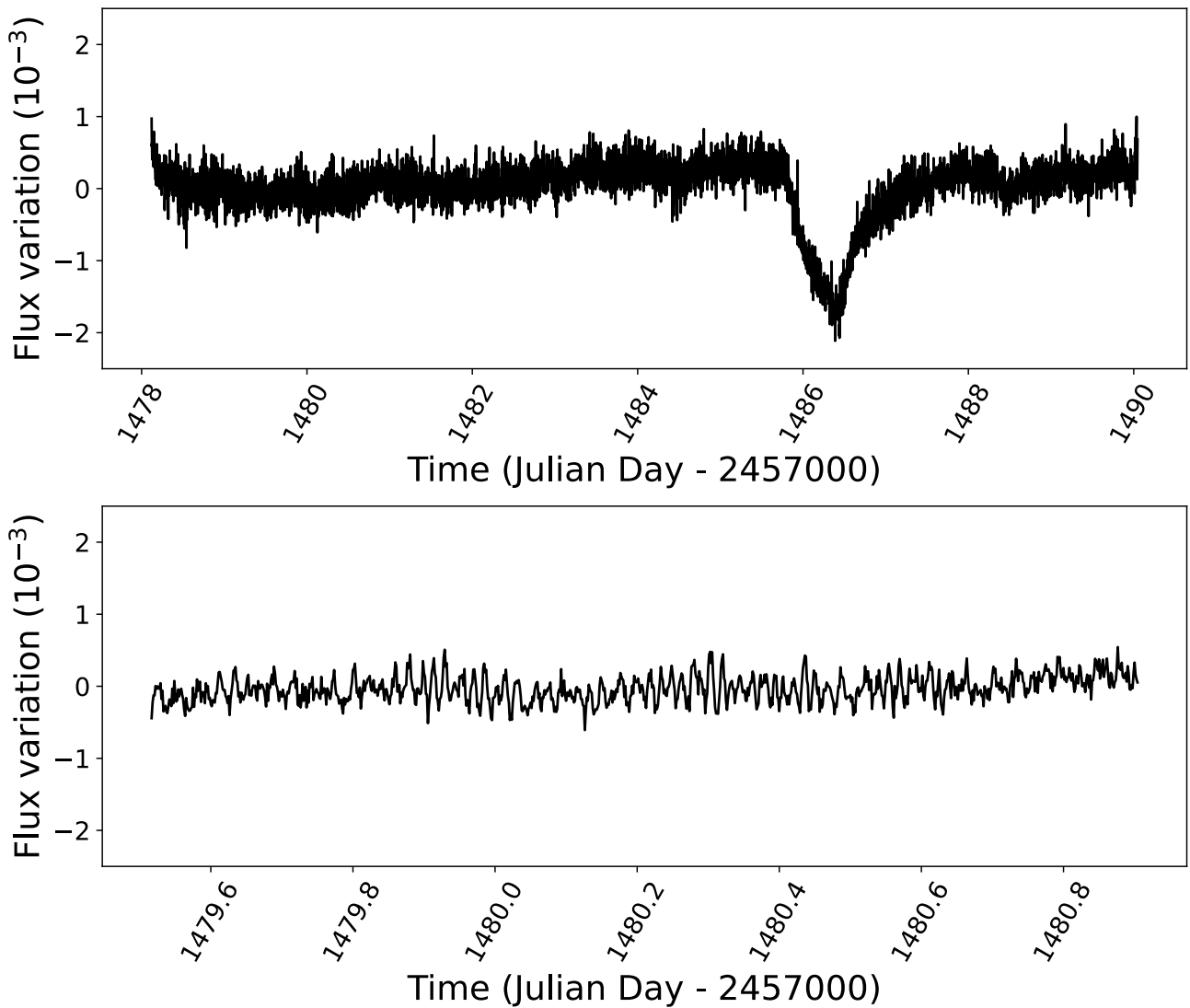}
      \caption{{\bf The \bp\ TESS light curve after removing the \ds\ photometric variations}. The plot is for the same time interval as in Extended Data Fig.~\ref{fig:delta scuti variations}. Here the photometric dip at Julian Day 2457000+1486 already identified in ref. \cite{Zieba_2019} is clearly visible. 
      }
        \label{fig:clean light curve}
  \end{figure}

  \begin{figure}[htp!]
   \centering
      \includegraphics[width=1\hsize]{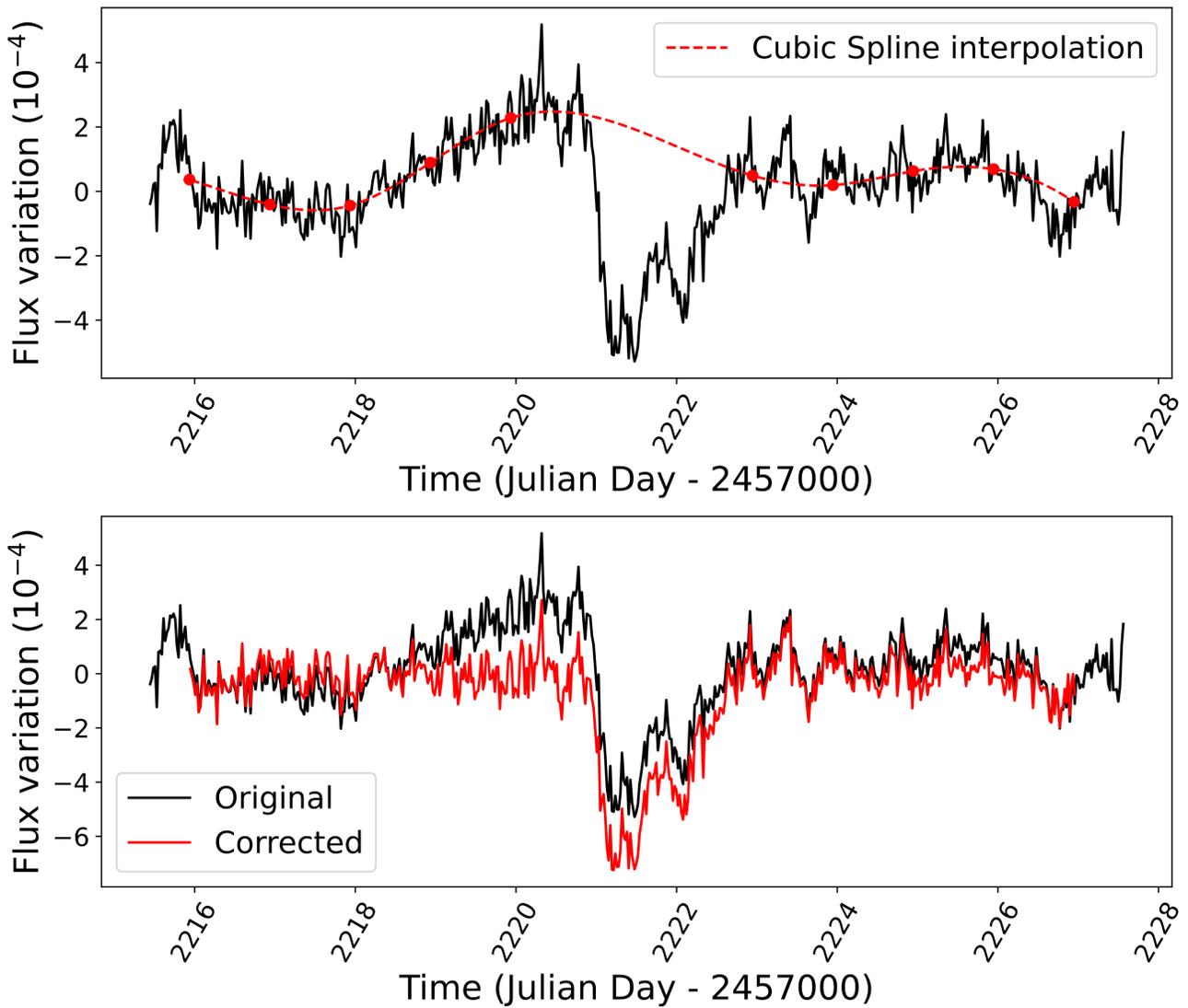}
      \caption{{\bf Illustration of the process to remove the slow time scale variations.} The top panel shows the data binned with a 1-day interval (red dots) and the cubic spline fitting the rebinned data (red dotted line). The bottom panel shows the original data (black line) and the final data after correction (red line).}
        \label{fig:final light curve}
  \end{figure}

  \begin{figure*}[htp!]
   \centering
   \includegraphics[width=0.85\textwidth]{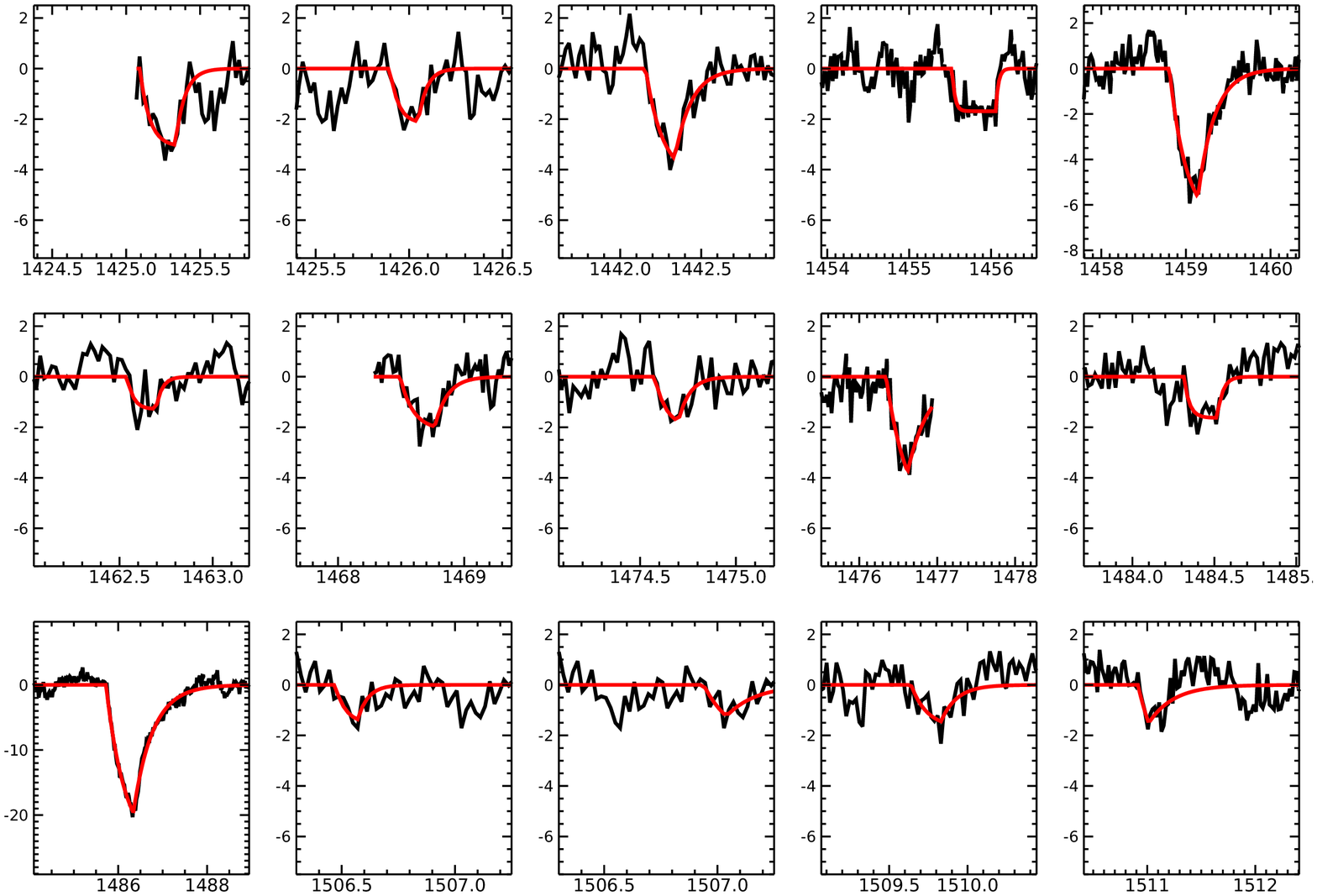}
   \includegraphics[width=0.85\textwidth]{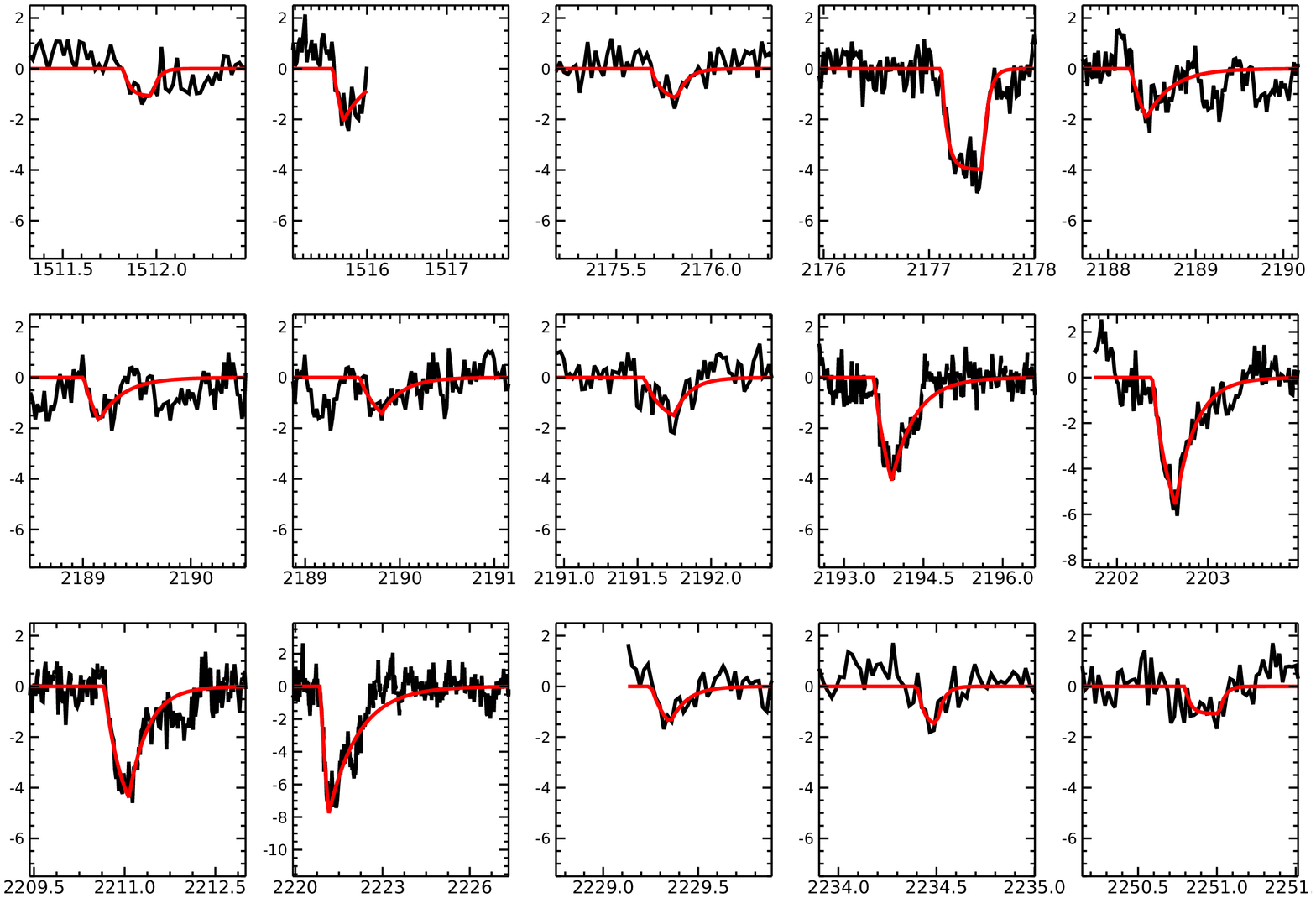}
   \caption{{\bf \bp\ photometric variations at the time of the 
   30~detected transits of exocomets.} 
   The time in the x-axis is given in units of Julian Day-2457000
   and the flux variations in the y-axis are relative to the mean flux and given in units of $10^{-4}$.  
   The red line represents the best fits with the 1-D model. 
}
        \label{fig:dips}
  \end{figure*}

\newpage

\section*{Extended Data Tables}

\setcounter{table}{0}
\renewcommand{\tablename}{Extended Data Table}

\begin{table}[ht!]
\caption{{\bf Log of TESS observations.}
The epochs of the data studied in ref.~\cite{Zieba_2019} are noted in bold font.}
        \label{tab:tess_data}
        \centering
        \begin{tabular}{cc}
        \hline
            Date begin  & Date end \\ 
            (JD-2457000) & (JD-2457000) \\ \hline
            {\bf 1411.0} & {\bf 1418.5} \\  
            {\bf 1425.1} & {\bf 1436.1} \\  
            {\bf 1438.0} & {\bf 1450.1} \\  
            {\bf 1451.6} & {\bf 1463.9} \\  
            {\bf 1468.3} & {\bf 1476.9} \\  
            {\bf 1478.1} & {\bf 1490.0} \\  
            {\bf 1491.6} & {\bf 1503.0} \\  
            {\bf 1504.7} & {\bf 1516.0} \\  
                 2174.2  &      2185.7 \\  
                 2187.2  &      2200.0 \\  
                 2201.7  &      2213.8 \\  
                 2215.4  &      2227.5 \\  
                 2229.1  &      2240.8 \\  
                 2243.0  &      2253.0 \\  
            \hline
            \noalign{\smallskip}
        \end{tabular}
    \end{table}

\begin{table*}[htb]
\caption{{\bf List of the detected exocomet transits.} 
The second to the fifth columns give the parameters of the best fits with the 1-D model. The last column gives the corresponding calculated absorption depths.}
\label{tab:List detections}
\begin{center}
\begin{tabular}{r|ccrrrc}
\hline
ID  & Time $t_0$ & $\Delta t$ & $\beta$ & $K$ & \multicolumn{2}{r}{Absorption depth}\\
    & (JD-2457000) & (days) & (days$^{-1}$)  & ($10^{-4}$) &
    \multicolumn{2}{c}{($10^{-4}$)}\\
\hline
  1 &   1425.09 &    0.24 &   15.44 &    3.10 &    3.02 & $\pm$   0.26 \\
  2 &   1425.90 &    0.15 &   22.11 &    2.17 &    2.10 & $\pm$   0.29 \\
  3 &   1442.16 &    0.18 &    9.76 &    4.32 &    3.56 & $\pm$   0.20 \\
  4 &   1455.52 &    0.53 &   30.00 &    1.69 &    1.69 & $\pm$   0.15 \\
  5 &   1458.80 &    0.34 &    5.02 &    6.93 &    5.65 & $\pm$   0.15 \\
  6 &   1462.54 &    0.15 &   30.00 &    1.29 &    1.27 & $\pm$   0.27 \\
  7 &   1468.49 &    0.27 &    9.85 &    2.11 &    1.97 & $\pm$   0.21 \\
  8 &   1474.58 &    0.12 &   16.79 &    2.02 &    1.77 & $\pm$   0.27 \\
  9 &   1476.34 &    0.28 &    3.62 &    5.94 &    3.76 & $\pm$   0.18 \\
 10 &   1484.31 &    0.20 &   30.00 &    1.63 &    1.63 & $\pm$   0.29 \\
 11 &   1485.72 &    0.62 &    2.30 &   25.91 &   19.63 & $\pm$   0.15 \\
 12 &   1506.47 &    0.10 &   21.60 &    1.54 &    1.36 & $\pm$   0.30 \\
 13 &   1506.94 &    0.10 &    7.59 &    2.33 &    1.24 & $\pm$   0.30 \\
 14 &   1509.64 &    0.19 &    9.83 &    1.71 &    1.45 & $\pm$   0.27 \\
 15 &   1510.91 &    0.10 &    4.31 &    4.32 &    1.51 & $\pm$   0.30 \\
 16 &   1511.82 &    0.15 &   30.00 &    1.08 &    1.07 & $\pm$   0.25 \\
 17 &   1515.57 &    0.13 &    2.90 &    6.46 &    2.06 & $\pm$   0.27 \\
 18 &   2175.68 &    0.14 &   17.82 &    1.26 &    1.15 & $\pm$   0.31 \\
 19 &   2177.12 &    0.39 &   18.45 &    3.98 &    3.97 & $\pm$   0.15 \\
 20 &   2188.26 &    0.18 &    3.47 &    4.15 &    1.95 & $\pm$   0.20 \\
 21 &   2189.01 &    0.14 &    4.41 &    3.69 &    1.71 & $\pm$   0.25 \\
 22 &   2189.57 &    0.23 &    4.47 &    2.16 &    1.40 & $\pm$   0.15 \\
 23 &   2191.55 &    0.20 &    9.01 &    1.78 &    1.49 & $\pm$   0.22 \\
 24 &   2193.56 &    0.34 &    2.21 &    7.69 &    4.09 & $\pm$   0.15 \\
 25 &   2202.39 &    0.26 &    4.45 &    8.21 &    5.59 & $\pm$   0.17 \\
 26 &   2210.66 &    0.41 &    3.11 &    6.10 &    4.39 & $\pm$   0.15 \\
 27 &   2220.84 &    0.31 &    0.97 &   30.12 &    7.78 & $\pm$   0.15 \\
 28 &   2229.25 &    0.10 &   11.22 &    2.13 &    1.44 & $\pm$   0.28 \\
 29 &   2234.40 &    0.10 &   30.00 &    1.57 &    1.50 & $\pm$   0.28 \\
 30 &   2250.80 &    0.22 &   28.59 &    1.08 &    1.08 & $\pm$   0.31 \\
\hline
\end{tabular}
\end{center}
\end{table*}

\end{document}